\documentclass[12pt]{article}

\usepackage{graphicx} 
\usepackage{amsmath, amssymb} 
\usepackage{float} 
\usepackage{hyperref} 
\usepackage{geometry} 
\usepackage{caption} 
\usepackage{booktabs} 
\usepackage{adjustbox} 
\usepackage{subfigure}
\usepackage{graphicx}
\usepackage{subcaption} 
\usepackage{float} 
\usepackage{geometry} 

\title{The physics of oscillating surfaces and sounds}
\author{Md Hossen Mondal\\
\textit{Department of Physics, Jadavpur University, West Bengal , India} \\ 
\and Ramkrishna Joshi\\
\textit{Department of Physics, Indian Institute of Technology Hyderabad , India}}
\date{\today}

\begin{document}

\maketitle

\begin{abstract}
The longitudinal oscillations of air columns composed of contractions and rarefactions make up sound. Sound amplification is widely used in medical, electronic and communication fields. A simplistic technique for producing and amplifying can be rewarding. In this study, we investigate a simplistic DIY speaker configuration that can be utilized for sound creation and modulation by implementing response of magnets and a solenoid to an oscillating input signal. We use steady state solution of forced simple harmonic oscillator with damping parameters to analyze our design and show its characteristic frequencies. We present an analytical way of obtaining optimal parameters of the setup to theoretically obtain experimental characteristic frequencies and provide an in-depth investigation of the setup.

\end{abstract}

\section{Introduction}
A speaker essentially is an electronic circuit that takes in an electric signal and generates sound as output. The mechanism underlying this process is essential to understand while studying sound. A traditional speaker operates by converting electrical signals into sound waves through a series of coordinated mechanisms. The core component is the voice coil, which is attached to a diaphragm (usually a cone). When an audio signal passes through the coil, it generates a varying magnetic field. This field interacts with the magnetic field of a fixed permanent magnet, causing the coil and attached diaphragm to vibrate. These vibrations create pressure waves in the surrounding air, producing sound. The frequency and amplitude of the electrical signal determine the pitch and loudness of the sound, respectively. The speaker enclosure plays a crucial role in enhancing sound quality by reducing interference and amplifying specific frequencies. This combination of electromagnetic and mechanical processes allows speakers to reproduce audio signals as audible sound. In this study, we demonstrate a simple DIY set up of a speaker with minimum component requirement. This setup can be generalized to produce sound from any surface.  

\begin{figure}
    \centering
    \includegraphics[width=0.7\linewidth]{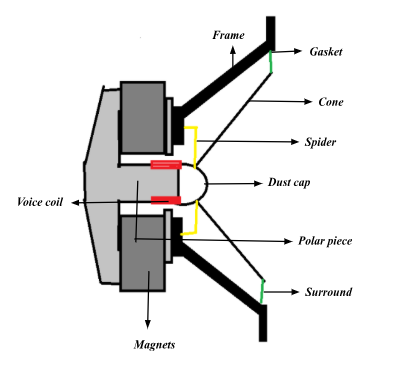}
    \caption{Assembly of a traditional speaker used to amplify and modulate the sound}
    \label{fig:enter-label}
\end{figure}

The speaker assembly consists of several key components, each playing a vital role in sound reproduction. All components of the speaker are mounted in a frame. The gasket prevents air leaks that may affect the quality of sound production. The cone vibrates to produce sound waves responding to the input electrical signals. Dust cap functions as a protective layer to keep out the dust and debris. voice coil, a wound copper or aluminum wire, moves within a magnetic field to drive the cone's motion. Voice coil form acts a bridge to transfer oscillations from the voice coil to the cone. Spider maintains the centered position of the voice coil in the magnetic field. The pole piece focuses the magnetic field around the voice coil. This field induces stronger oscillations in the voice coil thus generating sound effectively.

Any conducting wire creates a magnetic field around it, the direction of which can be deduced from Right Hand Thumb Rule (RHTR). A solenoid can be made by winding a conducting wire around a base. The nature of the magnetic field is well determined. The magnetic field inside is uniform throughout the region and is given by;\\

\begin{equation}
B = \mu_{0} n I  
\end{equation}

Where, B is the magnetic field inside a solenoid, $\mu_{0}$ is the permeability constant (of the core), n is the number of turns per unit length and I is the current passing through a solenoid. The magnetic field inside a solenoid can be derived using Ampère's law, which states:
\begin{equation}
\oint \vec{B} \cdot {\vec{l}} = \mu_0 I_{\text{enc}},
\end{equation}
where \( \vec{B} \) is the magnetic field, \( {\vec{l}} \) is the infinitesimal line element along the closed path of integration, \( \mu_0 \) is the permeability of free space, and \( I_{\text{enc}} \) is the current enclosed by the chosen path.

Consider an ideal solenoid with \( n \) turns per unit length and carrying a current \( I \). We assume the solenoid is infinitely long, so the magnetic field inside is uniform and parallel to the axis of the solenoid, while the field outside is negligible.

To apply Ampère's law, we choose a rectangular Amperian loop such that one side of the loop lies inside the solenoid, parallel to its axis and the opposite side lies outside, where the magnetic field is approximately zero.

For the portion of the loop inside the solenoid, Ampère's law simplifies to:
\begin{equation}
B \cdot l = \mu_0 I_{\text{enc}},
\end{equation}
where \( l \) is the length of the solenoid along the loop and $I_{\text{enc}} = n \cdot l \cdot I,$. The algebra follows to give, $B = \mu_0 n I$. Refer to \hyperlink{https://archive.org/details/davidgriffithselectrodynamics}{[5]} for detailed derivation.
Thus, the magnetic field inside an ideal solenoid is directly proportional to the current \( I \) and the turn density \( n \), and it is uniform along the axis of the solenoid.

\begin{figure}[H]
    \centering
    \begin{minipage}{0.58\textwidth}
        \centering
        \includegraphics[width=\linewidth]{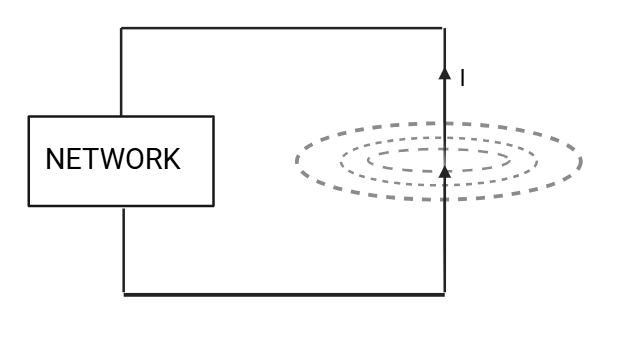} 
        \\(a)
    \end{minipage}
    \hfill
    \begin{minipage}{0.36\textwidth}
        \centering
        \includegraphics[width=\linewidth]{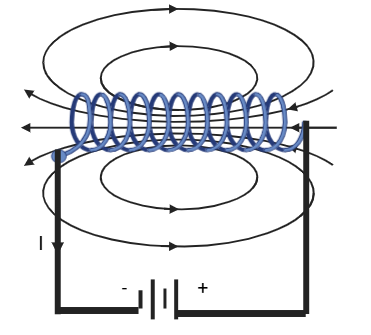} 
        \\(b)
    \end{minipage}

    \caption{Magnetic fields produced by (a) Straight current carrying wire and (b) Solenoid. The direction of field lines can be determined by using right hand thumb rule (RHTR).}
    \label{fig:side_by_side_minipage}
\end{figure}

A constant DC current supplied to the solenoid will generate a  constant magnetic field inside the solenoid. However, sending an AC will generate an oscillating magnetic field which is crucial to pressure waves of sound. 
The direction of the magnetic field inside the solenoid can be determined by the Right Hand Thumb Rule (RHTR). For  AC current, I = $I_{0}$ sin ($\omega$t), direction of magnetic field reverses every half current cycle. Due to this,if a permanent magnet is placed inside the solenoid then, for half of the cycle, the magnet will move in one direction and for other half cycle it will move in opposite direction leading to an oscillatory motion of the magnet. This motion can be readily seen by using Fleming's left hand rule. On bringing this magnet in contact with a surface, the surface will vibrate along with the magnet thus leading to generation of pressure waves of sound. This setup works as a simple speaker.

The project looks at the building and understanding of a speaker setup that can act as a loudspeaker. In this project we exploit the idea of oscillations of a magnet caused inside a solenoid due to an oscillating magnetic field. The setup primarily uses easily accessible equipment and electronics. Apart from the accessibility, the setup can be moved around to try and make different surfaces vibrate thus making them sing. As an example of this we have experimented with a white board, a table and a waste bin, and were successful in making them sing. This setup is simple,accessible and durable.

By changing the amplitude of pressure waves produced, it is possible to modulate the loudness of the generated sound. The intensity of the sound (I) is related to the amplitude as;\\
\begin{equation}
I \propto A^2 
\end{equation}

A simple derivation follows;
The intensity of a wave is defined as the energy transmitted per unit area per unit time. It can be expressed as:

\begin{equation}
I = \frac{P}{a},
\end{equation}

where \( I \) is the intensity, \( P \) is the power of the wave, and \( a \) is the area through which the wave propagates. For a sinusoidal wave, the displacement of the medium at position \( x \) and time \( t \) can be written as:
\begin{equation}
y(x,t) = A \sin(kx - \omega t),
\end{equation}
where \( A \) is the amplitude, \( k \) is the wave number, and \( \omega \) is the angular frequency.

The velocity of the medium particles is given by:
\begin{equation}
v(x,t) = \frac{\partial y}{\partial t} = -A \omega \cos(kx - \omega t).
\end{equation}

The kinetic energy per unit mass of the medium is proportional to the square of the velocity:
\begin{equation}
KE = \frac{1}{2} m v^2 = \frac{1}{2} m (A \omega)^2 \cos^2(kx - \omega t).
\end{equation}

Since the total energy in the wave is proportional to the kinetic energy, the energy carried by the wave is proportional to:
\begin{equation}
E \propto A^2 \omega^2.
\end{equation}
The intensity \( I \) of a wave is proportional to the power \( P \), which is the rate of energy transfer:
\begin{equation}
P \propto E \propto A^2 \omega^2.
\end{equation}

Thus, the intensity of the wave is:
\begin{equation}
I \propto A^2.
\end{equation}
Hence, we analyze the efficiency of the speaker by squaring the output peak to peak voltage generated by the response of the speaker. Independent variables of the experiment have to be optimized in such a way that we get a sufficiently large output amplitude, thus making the sound louder. 

\section{Literature Review}

The idea of making a speaker from minimalist resources
has fascinated many. As discussed in \hyperlink{https://www.scientificamerican.com/article/build-your-own-speaker/}{[1]} and \hyperlink{https://www.researchgate.net/publication/253610203_Magnetostatics_Analysis_Design_and_Construction_of_a_Loudspeaker}{[2]}, there are multiple simple ways to construct an electric speaker. We utilize cylindrical neodymium magnets to generate a response of the speaker for the given frequency. The magnetic field generated by cylindrical permanent magnets has been extensively analyzed. As illustrated in \hyperlink{https://www.researchgate.net/publication/29599218_Magnetic_Field_Created_by_Thin_Wall_Solenoids_and_Axially_Magnetized_Cylindrical_Permanent_Magnets}{[3]}, a cylindrical permanent magnet with uniform axial polarization \( \vec{J} \) produces the same magnetic field as a thin wall solenoid with equivalent dimensions when currents flow with a linear density \( \vec{K} \). This equivalence can also be modeled by a cylindrical current sheet. The relationship between the polarization \( \vec{J} \) and the linear current density \( \vec{K} \) is expressed as:

\
\begin{equation}
\vec{K} = \frac{\vec{J}}{\mu_0}
\end{equation}
\

where \( \mu_0 \) represents the permeability of free space.

Furthermore, the magnetic field can also be replicated using two charged planes situated at the top and bottom caps of the cylinder. These planes exhibit opposite surface charge densities \( +\sigma^* \) and \( -\sigma^* \), which are related to the polarization through:

\begin{equation}
\sigma^* = \vec{J} \cdot \vec{n}
\end{equation}
where \( \vec{n} \) is the unit vector normal to the surface.

For practical materials such as neodymium iron boron magnets, typical parameters include:
\[
J = 1.4 \, \text{T}, \quad \sigma^* = 1.4 \, \text{T}, \quad K = 1.11 \times 10^6 \, \text{A/m}.
\]

The system of electric speaker can be effectively analyzed as steady state oscillations of a forced harmonic oscillator (FHO). The generic solution for this steady-state oscillation can be readily derived from the general equation of motion.The equation of motion for a damped harmonic oscillator, influenced by an external time-dependent force \( F(t) = mf(t) \), is expressed as \hyperlink{https://web.mit.edu/8.Math/www/lectures/lec4/1.4.5.pdf}{[4]}:
\begin{equation}
\mathcal{L}[x(t)] \equiv \ddot{x} + \gamma \dot{x} + \omega_0^2 x = f(t),
\end{equation}
where \( \mathcal{L} \) denotes the differential operator, \( \gamma \) is the damping coefficient, and \( \omega_0 \) represents the natural frequency. To analyze the response of the system to a periodic driving force, the force is represented as \( f_\omega \cos(\omega t) \), where \( \omega \) is the driving frequency.
\begin{equation}
\therefore \ddot{x} + \gamma \dot{x} + \omega_0^2 x = \Re \big[ f_\omega e^{i \omega t} \big],
\end{equation}
allowing for an effective treatment of the problem in the frequency domain. Damping parameter plays a crucial role in determining the amplitude of the steady state oscillation. Theoretically, the resonant frequency of the setup is highly sensitive to the changes in damping parameter. A zero damping coefficient accounts for an infinite oscillation amplitude at resonant frequency and hence infinite intensity. The material of the sound box prepared can be characterized in this experiment by accounting for varying damping coefficients $\gamma$. Sound transmission loss (STL) quantifies sound insulation properties of a material.

\begin{equation}
\text{STL} = 10 \log_{10} \left( \frac{W_i}{W_t} \right) = 10 \log_{10} \left( \frac{1}{\tau} \right)
\end{equation}

where \(W_i\) is incident sound power and \(W_t\) is transmitted sound power in W, and \(\tau = \frac{W_t}{W_i}\) is Sound transmission coefficient.

STL curve for neat rHDPE consists of four main regions, namely; the stiffness-controlled region in 63-100 Hz, the resonance-controlled region in 125-500 Hz, mass-controlled region in 630-3150 Hz, coincidence-controlled region for $> 4000 Hz$ region \hyperlink{https://www.ioa.org.uk/system/files/proceedings/j_mago_sunali_a_negi_s_fatima_sound_insulation_property_of_recycled_high-density_polyethylene_waste_jute_fabric_composites.pdf}{[7]}.In this experiment we mainly operate in the resonance-controlled region where STL depends upon material damping. Hence accounting for an accurate $\gamma$ plays a crucial role in the analysis.

\section{Methodology}

The setup of the experiment includes a laboratory-made solenoid, an amplifier board, neodymium magnets, a breadboard-based microphone circuit, a sound box, and holding stands. The solenoid core was a hollow PVC cylinder of length \( (8.3 \pm 0.1) \, \text{cm} \) and diameter \( (2.4 \pm 0.05) \, \text{cm} \). A 3 mm gauge copper wire was wound for \( 154 \pm 5 \) turns. A 5L water can cut in half was used as a sound box. A DC12-24V TPA3116D2 Subwoofer Amplifier Board was utilized. Six-seven neodymium magnets were used for a high-intensity output from the sound box.
\vspace{-0.5mm}
\begin{figure}[H]
    \centering
    \begin{minipage}{0.45\textwidth}
        \centering
        \includegraphics[width=\linewidth]{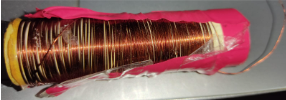} 
        \\(a)
    \end{minipage}
    \hfill
    \begin{minipage}{0.35\textwidth}
        \centering
        \includegraphics[width=\linewidth]{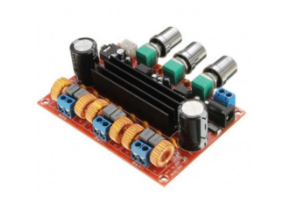} 
        \\(b)
    \end{minipage}

    \vskip\baselineskip 

    \begin{minipage}{0.45\textwidth}
        \centering
        \includegraphics[width=\linewidth]{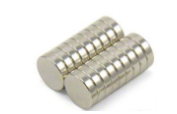} 
        \\(c)
    \end{minipage}
    \hfill
    \begin{minipage}{0.30\textwidth}
        \centering
        \includegraphics[width=\linewidth]{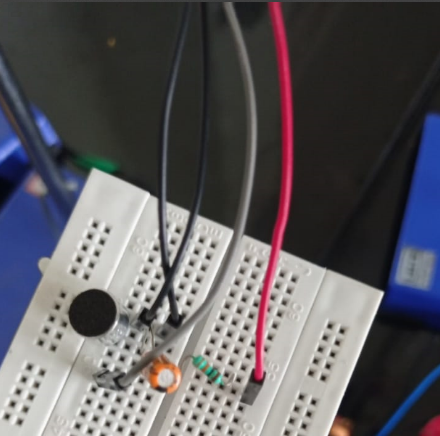} 
        \\(d)
    \end{minipage}

    \caption{Components utilized to prepare the sound box experimental setup. (a) Solenoid, (b) DC12-24V TPA3116D2 Subwoofer Amplifier Board, (c) Neodymium Magnets, (d) Breadboard based microphone circuit.}
    \label{fig:2x2_grid}
\end{figure}

The general methodology of experiment is given below:

\begin{itemize}
\item Choose the surface of oscillation and attach 5-6 neodymium magnets to that surface.
\item Wind up a chosen core hollow cylinder with the copper wire to make a solenoid and attach the two ends of the copper wire to the amplifier. In this experiment we use DC12-24V TPA3116D subwoofer amplifier board.
\item Attach the solenoid onto the surface such that the neodymium magnets lie inside the hollow core cylinder. Preferably, magnets should lie exactly at the center of the hollow cylinder rod for maximum efficiency of the speaker.
\item Attach the amplifier chord to a sound source and play the sound.
\end{itemize}

An iron core inserted in the solenoid highly enhances the magnetic field of the solenoid, in turn enhancing the output intensity of the sound. Thus, speaker setup can be made more efficient by experimenting with setup parameters. To record the output intensity from the speaker we prepared a electronic microphone circuit.

\begin{figure}[H]
    \centering
    \includegraphics[width=0.5\linewidth]{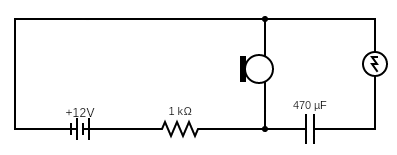}
    \caption{Electronic circuit of microphone built by using a $1 k\Omega$ resistor and $470 \mu F$ capacitor. A 12 V supply is connected.}
\end{figure}

This circuit is a simple microphone amplifier that uses a capacitive microphone to capture sound. The microphone generates a small AC voltage corresponding to sound vibrations. A resistor (1 k$\Omega$) is connected to provide biasing and current for the microphone to operate. A 470 $\mu$F capacitor is used as a coupling capacitor to block the DC component of the microphone signal. The AC signal from the microphone passes through the capacitor and reaches the connected load, which could be a speaker or further amplification stages. The +12V power supply provides the necessary voltage for the circuit to function properly. The capacitor isolates the DC bias voltage and allows only the AC signal to pass. The resistor plays a crucial role in setting the operating point of the microphone. This simple circuit can amplify sound signals for basic audio applications and is suitable for low-power audio input systems, making it ideal for DIY audio projects.

\begin{figure}[H]
    \centering
    \begin{minipage}{0.45\textwidth}
        \centering
        \includegraphics[width=\linewidth]{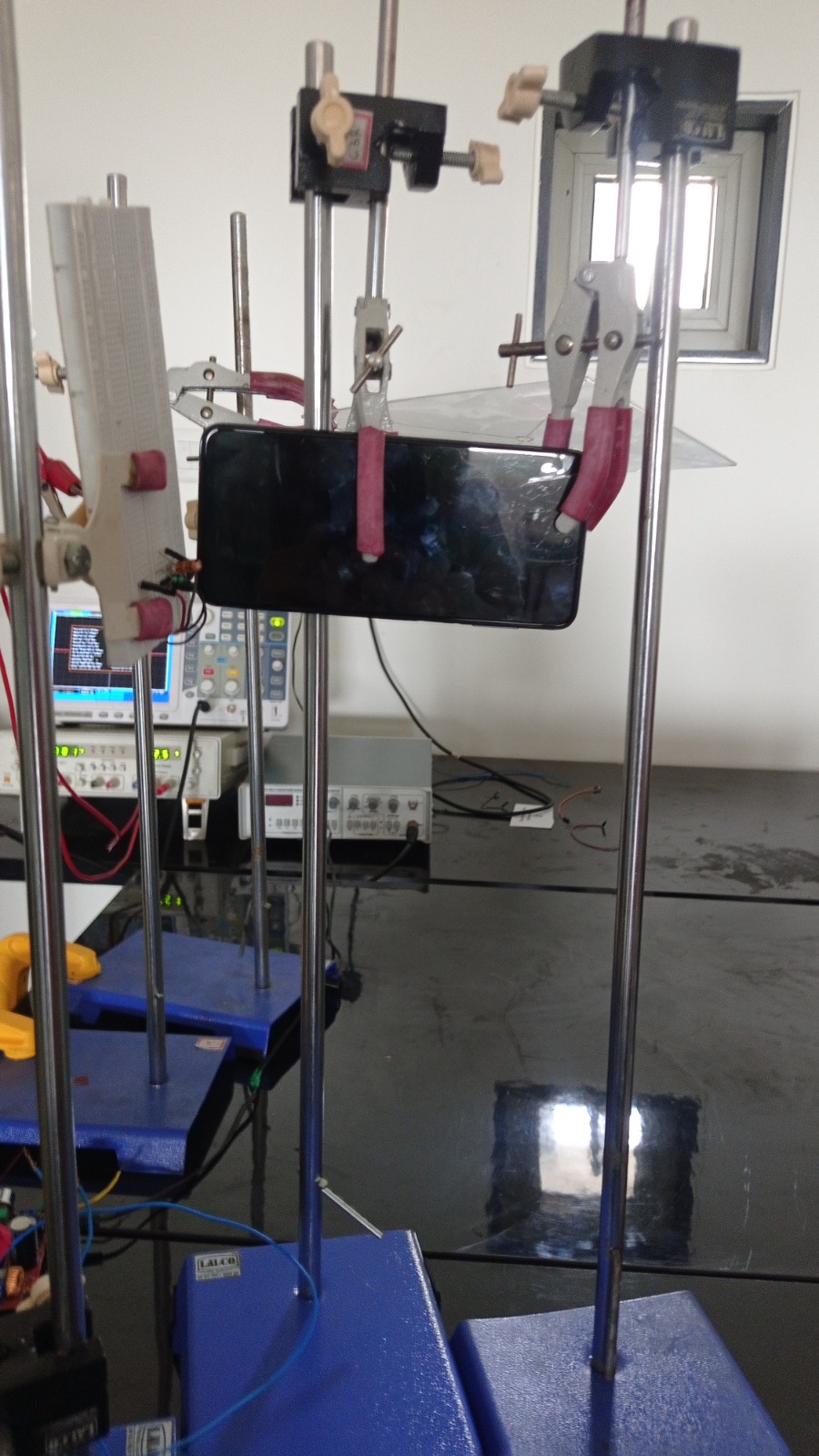} 
        \\(a)
    \end{minipage}
    \hfill
    \begin{minipage}{0.45\textwidth}
        \centering
        \includegraphics[width=\linewidth]{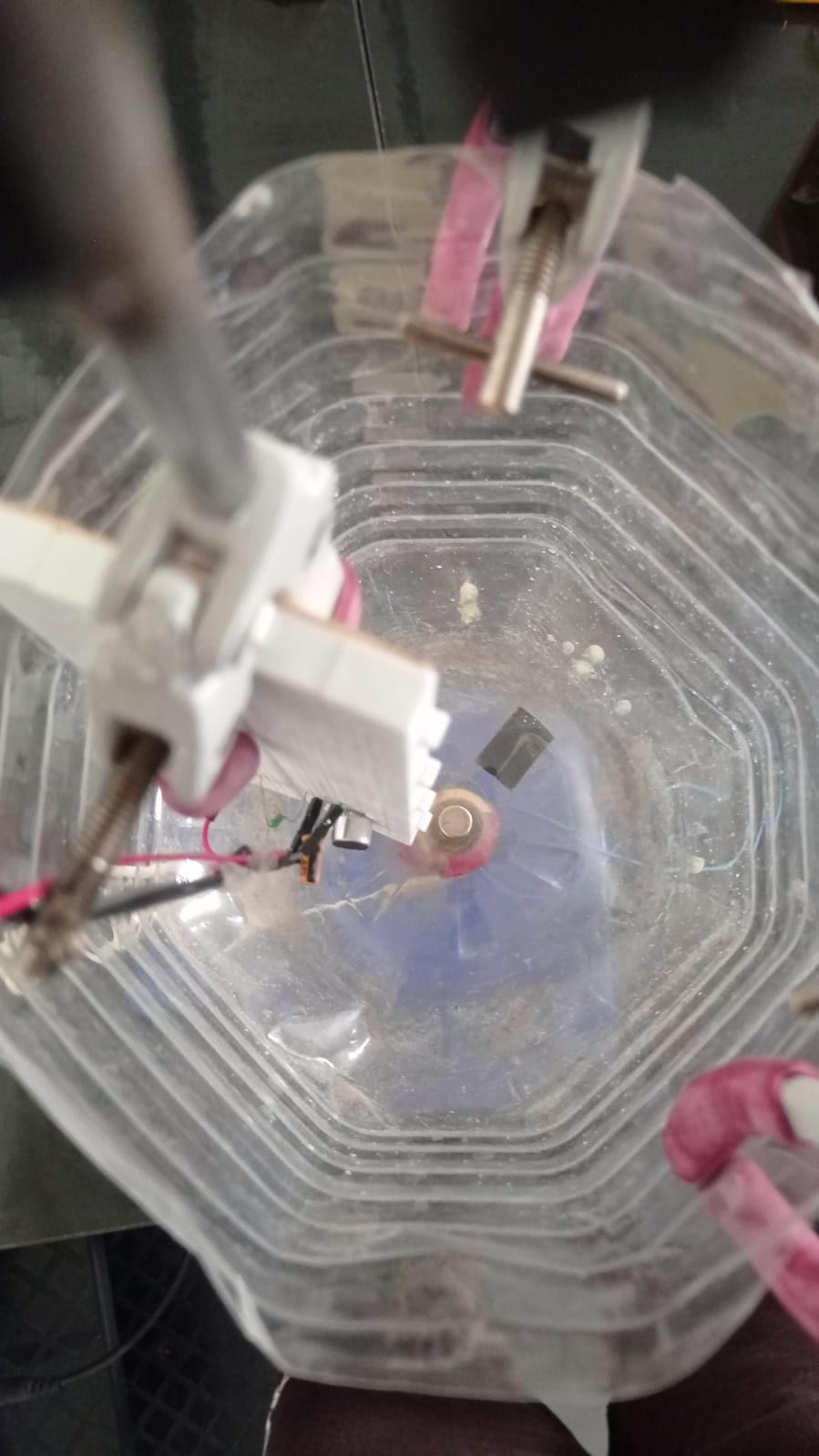} 
        \\(b)
    \end{minipage}

    \vskip\baselineskip 

    \begin{minipage}{0.65\textwidth}
        \centering
        \includegraphics[width=\linewidth]{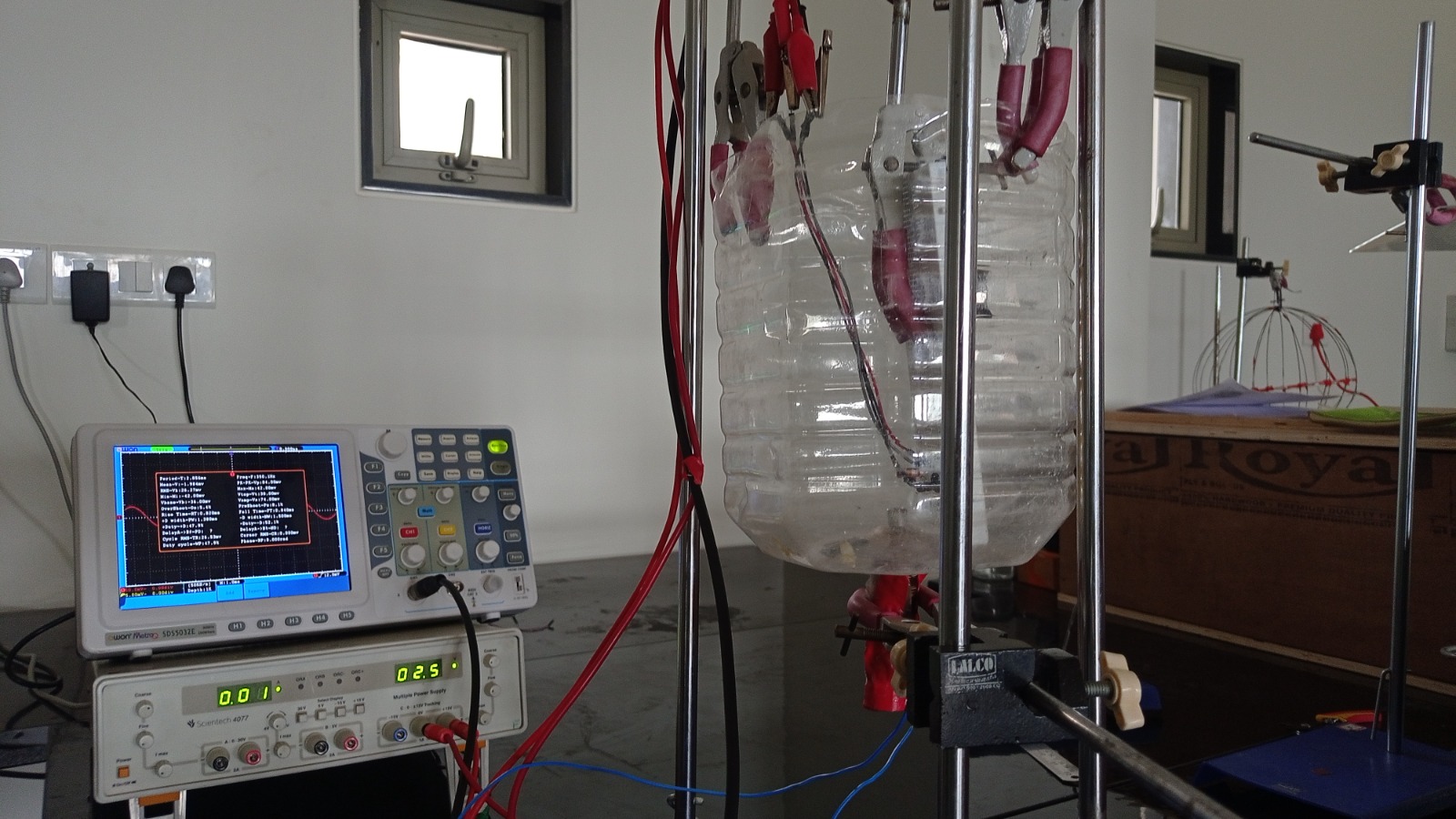} 
        \\(c)
    \end{minipage}

    \caption{(a) Phone setup side view, (b) Speaker setup top view and (c) Speaker setup side view. Output intensity was recorded by using the laboratory made microphone circuit.}
    \label{fig:2x2_grid}
\end{figure}

\begin{figure}[H]
    \centering
    \begin{minipage}{0.55\textwidth}
        \centering
        \includegraphics[width=\linewidth]{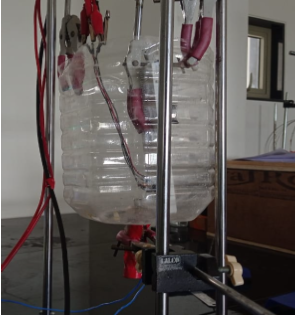} 
        \\(a)
    \end{minipage}
    \hfill
    \begin{minipage}{0.55\textwidth}
        \centering
        \includegraphics[width=\linewidth]{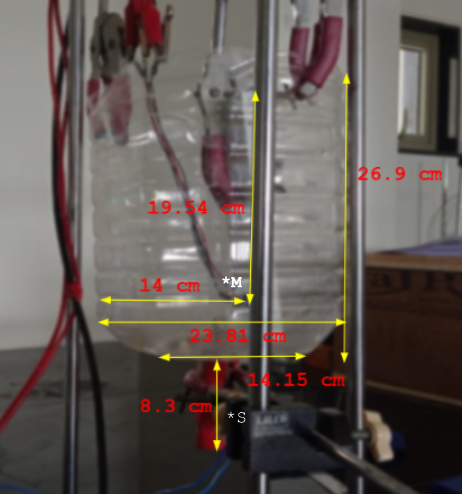} 
        \\(b)
    \end{minipage}
    
    \caption{The speaker setup with a a solenoid ($*S$) and magnets secured to the sound box and breadboard based microphone placed ($*M$) inside the sound box to record the sound output.}
    \label{fig:side_by_side_minipage}
\end{figure}

\begin{figure}[H]
    \centering
    \includegraphics[width=0.8\linewidth]{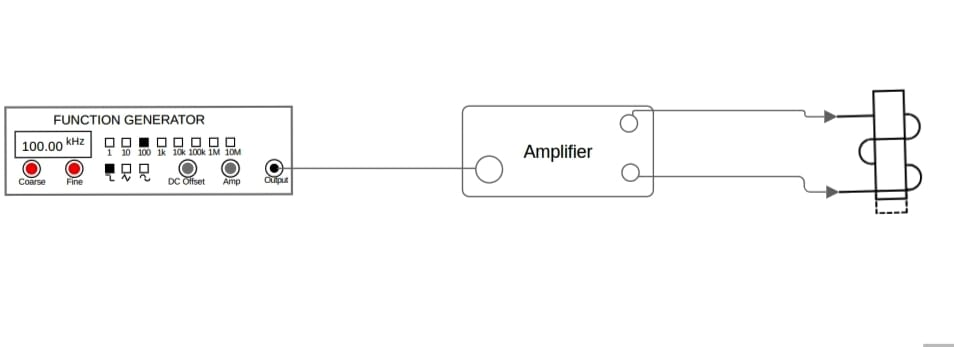}
    \caption{Schematic representation of setup}
    \label{fig:enter-label}
\end{figure}

A setup was prepared with phone and sound box and output peak to peak amplitude was recoded for a set of five frequencies for each setup. Copper wire is tightly wound around the solenoid core. Stripped wires were used to ensure proper connections. Solenoid was connected to the amplifier. Magnets were inside the solenoid which was then attached to the surface. On passing the modulated electric signal through the amplifier, vibrational modes of solenoid were excited leading to vibration of the surface. This, in turn produces pressure waves of sound.

\section{Results and Discussion}
\subsection{Data Comparison}

With the setup, we were able to study the response of a
speaker box and a white board to 5 different frequencies
of 150Hz,200Hz,250Hz,300Hz and 350Hz. This is a
comprehensive way of understanding the response of the
sound box and the white board. One in principle can
study response over a continuous range of frequencies.
Sound box Using a physics tool box application for the
android phone, we provided 5 discrete frequencies as
inputs to the mounted sound box. Each frequency has a
different vibration pattern/frequency for the
magnets,thus leading to a different response from the
sound box. keeping the master volume and the
subwoofer volume on maximum,accordingly the stereo
volume was adjusted to get the maximum output from
the sound box. A simple breadboard circuit was made
to act as a speaker. By introducing the speaker close
enough to the sound box, output caught by the speaker
was transferred to an oscilloscope to get the waveform
of the output. The output amplitude shown by the
oscilloscope was used as a measure of the intensity of
the sound produced by the sound box. The same
procedure was used to analyze the output from the
phone speaker. This provides a comparison between the
output of the sound box and the output of the phone
speaker. This can be used to determine the efficiency of
the setup; sound box in particular.
The speaker setup's response was compared to that of a phone speaker across various frequencies. The following table summarizes the results:

\begin{table}[H]
    \centering
    \begin{adjustbox}{max width=\textwidth}
    \begin{tabular}{|c|c|c|c|}
    \hline
    Frequency (Hz) & Pk-Pk (mV) & RMS (mV) & Intensity (mV$^2$) \\ \hline
    \multicolumn{4}{|c|}{\textbf{Speaker Output}} \\ \hline
    150 & 54 & 14.16 & 2916 \\ \hline
    200 & 48 & 16.21 & 2304 \\ \hline
    250 & 86 & 28.15 & 7396 \\ \hline
    300 & 288 & 99.27 & 82944 \\ \hline
    350 & 84 & 26.27 & 7056 \\ \hline
    \multicolumn{4}{|c|}{\textbf{Phone Output}} \\ \hline
    150 & 52 & 14.53 & 2704 \\ \hline
    200 & 72 & 23.44 & 5184 \\ \hline
    250 & 126 & 41.13 & 15876 \\ \hline
    300 & 183 & 67.34 & 33489 \\ \hline
    350 & 286 & 91.00 & 81796 \\ \hline
    \end{tabular}
    \end{adjustbox}
    \caption{Comparison of RMS and Peak-Peak outputs for the Speaker and Phone across five input frequencies ranging from 150 Hz to 350 Hz at regular intervals of 50 Hz.}
    \label{tab:comparison}
\end{table}

Peak-Peak response values were considered as the the amplitudes of the oscillatory waveform. Superposed plot of speaker output is compared with respect to the plot of phone output to study the resonant behavior of the sound box.

\begin{figure}[H]
    \centering
    \begin{minipage}{0.45\textwidth}
        \centering
        \includegraphics[width=\linewidth]{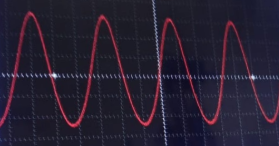} 
        \\(a)
    \end{minipage}
    \hfill
    \begin{minipage}{0.45\textwidth}
        \centering
        \includegraphics[width=\linewidth]{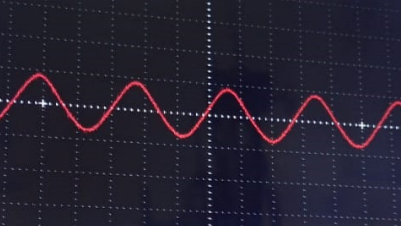} 
        \\(b)
    \end{minipage}

    \vskip\baselineskip 

    \begin{minipage}{0.45\textwidth}
        \centering
        \includegraphics[width=\linewidth]{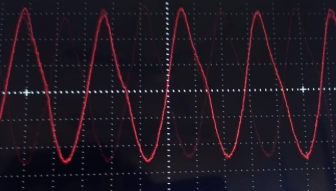} 
        \\(c)
    \end{minipage}
    \hfill
    \begin{minipage}{0.45\textwidth}
        \centering
        \includegraphics[width=\linewidth]{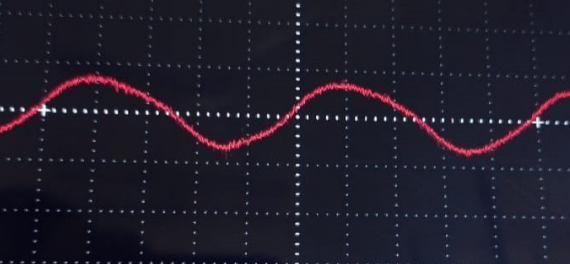} 
        \\(d)
    \end{minipage}

    \caption{Oscilloscope output waveforms obtained for (a) Speaker setup at 300 Hz, (b) Speaker setup at 350 Hz, (c) Phone setup at 350 Hz and (d) Phone setup at 250 Hz. Using these waveforms, sinusoidal response of the speaker and phone is reconstructed.}
    \label{fig:2x2_grid}
\end{figure}

\begin{figure}[H]
    \centering

    \begin{minipage}[t]{0.42\textwidth}
        \centering
        \includegraphics[width=\textwidth]{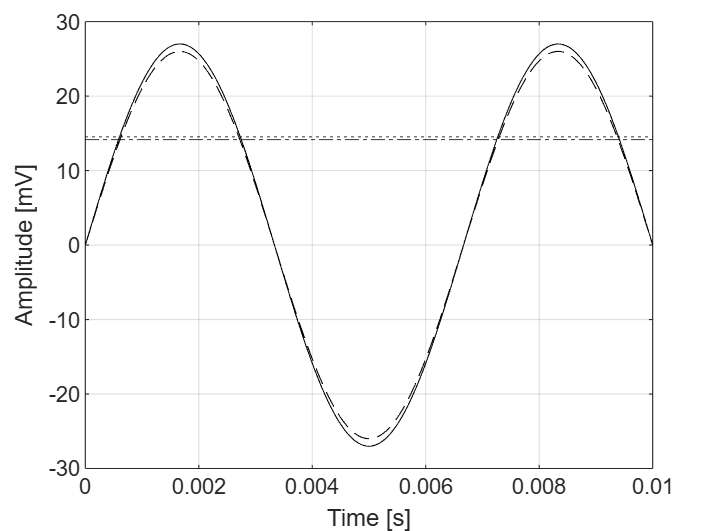}
        \vspace{0.1cm}
        \caption*{(a)}
    \end{minipage}
    \hfill
    \begin{minipage}[t]{0.42\textwidth}
        \centering
        \includegraphics[width=\textwidth]{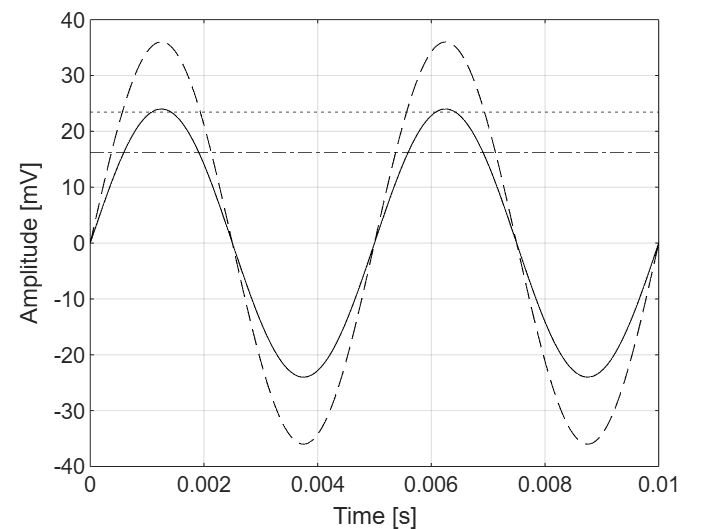}
        \vspace{0.1cm}
        \caption*{(b)}
    \end{minipage}

    \vspace{0.2cm} 
    \begin{minipage}[t]{0.42\textwidth}
        \centering
        \includegraphics[width=\textwidth]{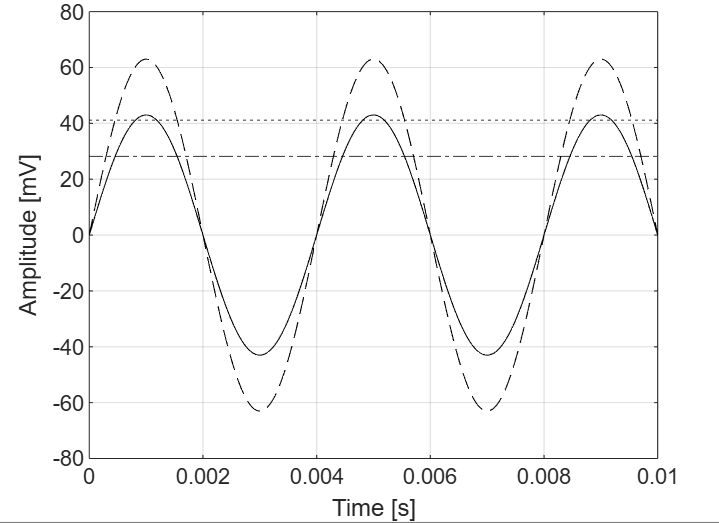}
        \vspace{0.1cm}
        \caption*{(c)}
    \end{minipage}
    \hfill
    \begin{minipage}[t]{0.42\textwidth}
        \centering
        \includegraphics[width=\textwidth]{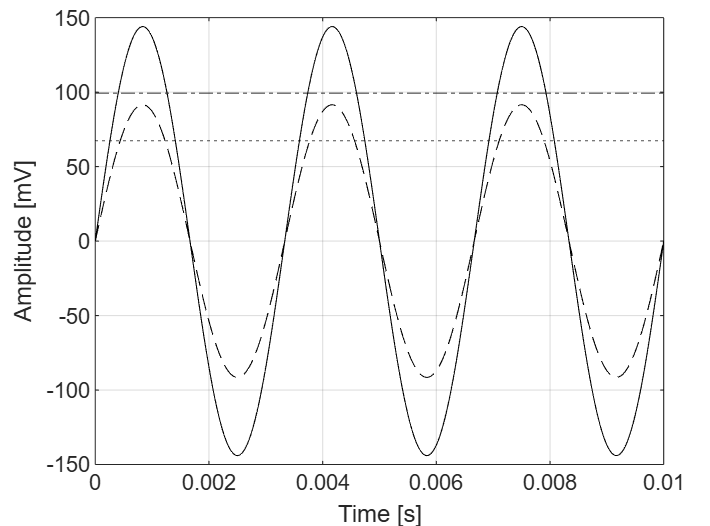}
        \vspace{0.1cm}
        \caption*{(d)}
    \end{minipage}

    \vspace{0.2cm} 

    \begin{minipage}[t]{0.42\textwidth}
        \centering
        \includegraphics[width=\textwidth]{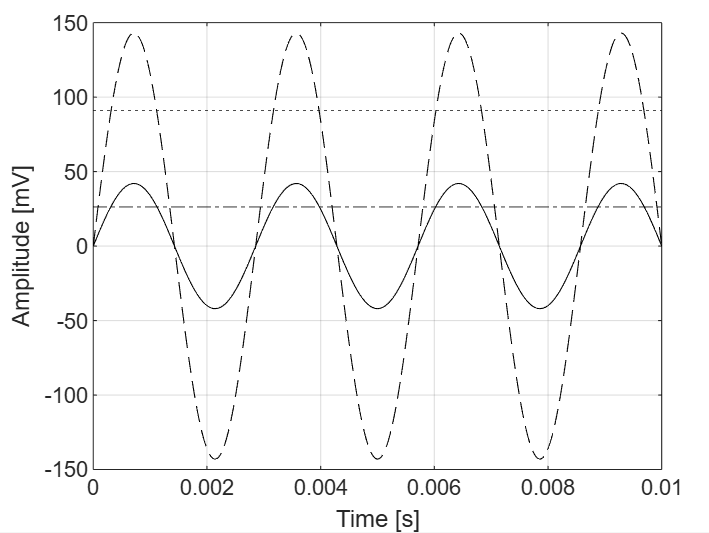}
        \vspace{0.1cm}
        \caption*{(e)}
    \end{minipage}
\caption{Reconstructed Peak-Peak sinusoidal voltage response of Speaker (Solid line) and Phone (Dashed line). RMS output voltages of speaker(Center-line) and Phone (Dotted line) are represented as lines parallel to time axis.}
\end{figure}

Intensity of output is defined as;
\begin{equation}
I_S \propto f_S^2
\end{equation}

\begin{equation}
I_P \propto f_P^2
\end{equation}

where subscripts S and P represent Speaker and Phone respectively. Intensity plot clearly highlights the resonant behavior of the speaker (sound box) setup.

\begin{figure}[H]
    \centering
    \includegraphics[width=0.9\textwidth]{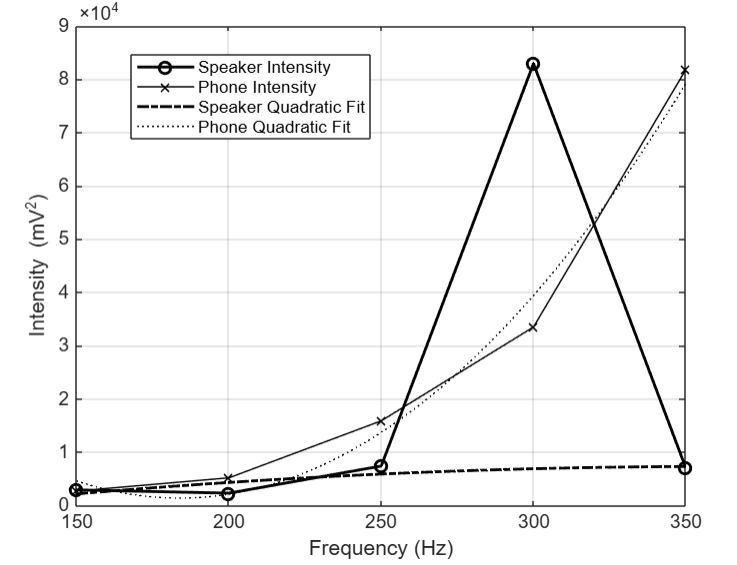} 
    \caption{Comparison plot showing speaker and phone output intensities.}
    \label{fig:intensity_comparison}
\end{figure}

The quadratic fitting functions used for the speaker and phone outputs are:
\begin{equation}
I_S = -0.1154 f_S^2 + 83.4156 f_S - 7753.9636
\end{equation}

\begin{equation}
I_P = 2.8164 f_P^2 - 1035.2363 f_P + 96509.9429
\end{equation}

\subsection{Heat maps for intensity fall at different frequencies}
Simulated heat maps illustrate the intensity distribution on a unit circular base for different frequencies. Actual sound box circular base radius of 7.75 cm is unitized to a circular base. In terms of experimental parameters, 1 m radius on a unit disc corresponds to the circular 7.75 cm radius. This Unit disc: Experimental disc conversion ratio is $\approx$14.13.

\begin{figure}[H]
    \centering
    \begin{minipage}{0.6\textwidth}
        \centering
        \includegraphics[width=\textwidth]{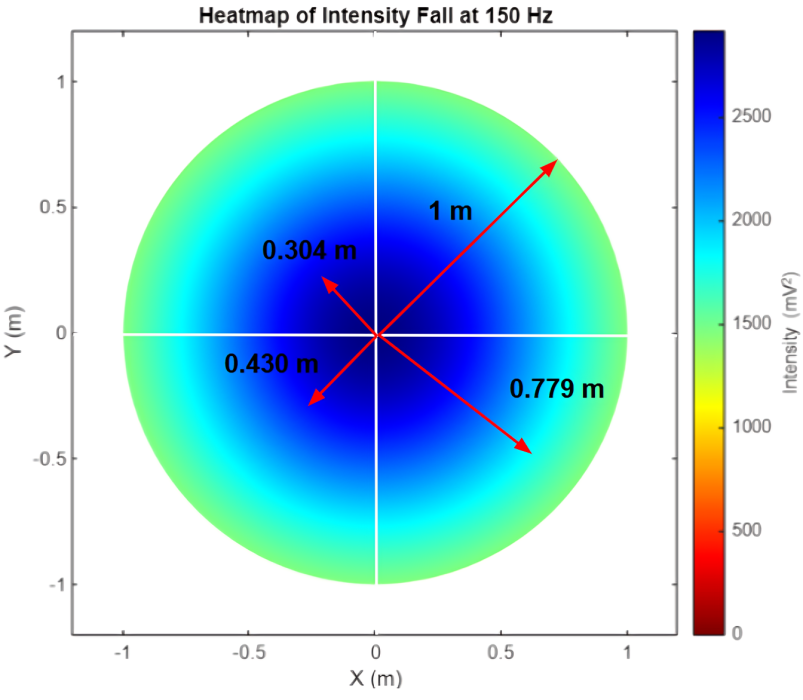} 
        \caption*{(a)}
    \end{minipage}
    \vspace{0.5cm}
    \begin{minipage}{0.6\textwidth}
        \centering
        \includegraphics[width=\textwidth]{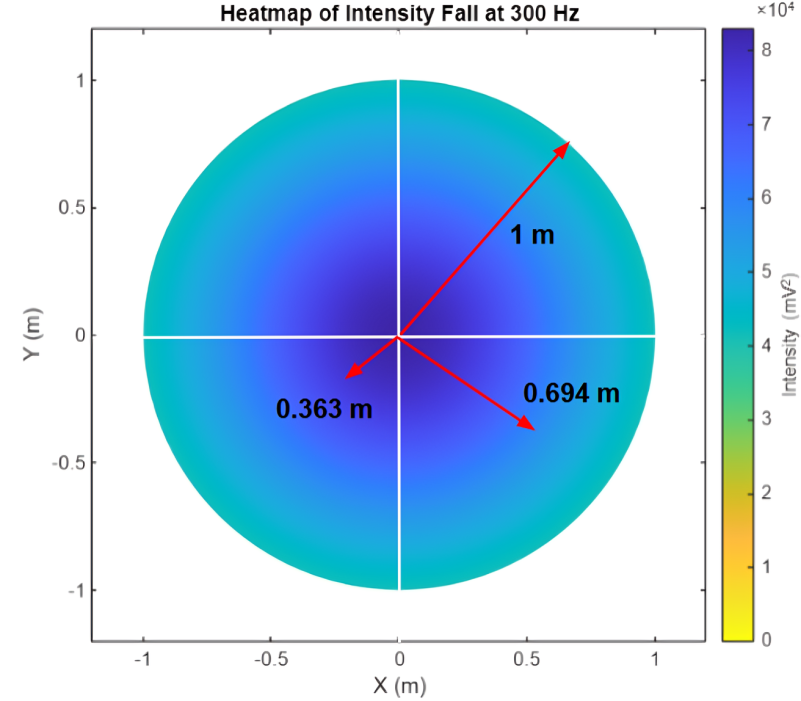} 
        \caption*{(b)}
    \end{minipage}
    \caption{Intensity fall heat maps for (a) 150 Hz and (b) 300 Hz. Intersection point of white cross represents a point source of vibrations. Red arrows represent directional vectors associated with intensity fall annular regions symmetric across the point source.}
    \label{fig:heatmaps}
\end{figure}

The model assumes that the intensity diminishes as the radial distance from the source increases due to spatial spreading. Intensity is assumed to decay exponentially over time due to medium absorption and energy losses.

Intensity $I$ as a function of radial distance $r$ and time $t$ is expressed as:
\begin{equation}
I(r, t) = \frac{I_0}{1 + r^2} \cdot \exp(-\lambda t),
\label{eq:intensity_model}
\end{equation}
where $I_0$ is the initial (central) intensity at the source, $r$ is the radial distance from the source in the plane of the speaker's base, $\lambda$ is the decay constant that governs the rate of intensity loss over time and $t$ is the time.

For the intensity fall model, inverse-square law accounts for the natural spreading of energy as a wave propagates radially outward from a source. The exponential decay term, $\exp(-\lambda t)$, represents the absorption and attenuation effects within the medium as energy is gradually lost over time. The model captures both spatial and temporal intensity attenuation. The term $(1 + r^2)$ is introduced instead of $r^2$ to avoid singularities at $r = 0$ and ensure mathematical stability, particularly in numerical computations. The source is considered to be a point source situated at $r = 0$ corresponding to $I_0$ intensity at the centre.

By applying this model for 150 Hz and 300 Hz, we predict the value of decay constant. At r=7.75 cm, the intensity value stands at $\approx 1500 mV^2$ for 150 Hz and $\approx 40000 mV^2$ for 300 Hz. Longitudinal sound speed in HDPE is $2401 \pm3\ m/s$\hyperlink{https://www.sciencedirect.com/science/article/pii/S0032386120309897}{[6]}. Hence sound travels the distance of 0.0775 m in T=$3.22 \times 10^ {-5}\ s$

\begin{equation}
1500\ mV^2 = \frac{2916\ mV^2}{1 + (0.0775)^2} \cdot \exp(-\lambda_{150Hz} T),
\label{eq:intensity_model}
\end{equation}

On substituting T; we obtain that $\lambda_{150Hz}= 2.04 \times 10^4\  s^-1$

Following similar analysis for 300 Hz frequency;

\begin{equation}
40000\ mV^2 = \frac{82944\ mV^2}{1 + (0.0775)^2} \cdot \exp(-\lambda_{300Hz} T),
\label{eq:intensity_model}
\end{equation}

we obtain $\lambda_{300Hz}= 2.24\times 10^4\  s^-1$. The average decay constant is found to be $2.14\times 10^4\  s^-1$.

\section{Theoretical model}

We analyze this system by considering steady state motion of a forced damped harmonic oscillator. Equations of motion for the system are:

\begin{equation}
m \left( \frac{d^2 y}{dt^2} \right) + b \left( \frac{dy}{dt} \right) = F_0 \cos(\omega t) - ky
\end{equation}

Where \( m \) is the mass of the system, \( b \) is the damping coefficient, \( k \) is the spring constant, \( F_0 \) is the driving force amplitude, \( \omega \) is the driving angular frequency. The steady state solution to this system is,

\begin{equation}
x(t) = x_0(\omega) \cos(\omega t + \varphi)
\end{equation}

With the oscillation amplitude given by,

\begin{equation}
x_0(\omega) = \frac{(F_0 / m)}{\sqrt{\left( \frac{b}{m} \right)^2 \omega^2 + ( \omega_0^2 - \omega^2 )^2}}
\end{equation}

And the phase shift is given by:

\begin{equation}
\varphi(\omega) = \tan^{-1} \left( \frac{(b \omega / m)}{\omega_0^2 - \omega^2} \right)
\end{equation}

Where \( \omega_0 = \sqrt{k / m} \) is the natural frequency of the system.

\subsection{Solution of the Damped Forced Harmonic Oscillator}

The equation of motion for the system is given by:
\begin{equation}
    m \frac{d^2y}{dt^2} + b \frac{dy}{dt} + ky = F_0 \cos(\omega t).
\end{equation}

Dividing through by \(m\), we obtain:
\begin{equation}
    \frac{d^2y}{dt^2} + \frac{b}{m} \frac{dy}{dt} + \frac{k}{m}y = \frac{F_0}{m} \cos(\omega t),
\end{equation}
where \(\gamma = \frac{b}{m}\), \(\omega_0^2 = \frac{k}{m}\), and \(f(t) = \frac{F_0}{m} \cos(\omega t)\). The equation becomes:
\begin{equation}
    \frac{d^2y}{dt^2} + \gamma \frac{dy}{dt} + \omega_0^2 y = f(t).
\end{equation}

The general solution is:
\begin{equation}
    y(t) = y_h(t) + y_p(t),
\end{equation}
where \(y_h(t)\) is the solution of the homogeneous equation and \(y_p(t)\) is a particular solution of the inhomogeneous equation.

The homogeneous equation is:
\begin{equation}
    \frac{d^2y}{dt^2} + \gamma \frac{dy}{dt} + \omega_0^2 y = 0.
\end{equation}
The characteristic equation is:
\begin{equation}
    \lambda^2 + \gamma \lambda + \omega_0^2 = 0,
\end{equation}
with roots:
\begin{equation}
    \lambda = -\frac{\gamma}{2} \pm \sqrt{\left(\frac{\gamma}{2}\right)^2 - \omega_0^2}.
\end{equation}
For \(\gamma^2 < 4\omega_0^2\), the solution is:
\begin{equation}
    y_h(t) = e^{-\gamma t / 2} \left(C_1 \cos(\omega_d t) + C_2 \sin(\omega_d t)\right),
\end{equation}
where \(\omega_d = \sqrt{\omega_0^2 - (\gamma/2)^2}\).

For the particular solution, we assume:
\begin{equation}
    y_p(t) = A \cos(\omega t) + B \sin(\omega t).
\end{equation}
Substituting into the equation, we find:
\begin{equation}
    A = \frac{F_0 (\omega_0^2 - \omega^2)}{m \left[(\omega_0^2 - \omega^2)^2 + (\gamma \omega)^2\right]},
    \quad
    B = \frac{F_0 \gamma \omega}{m \left[(\omega_0^2 - \omega^2)^2 + (\gamma \omega)^2\right]}.
\end{equation}
Thus, the particular solution is:
\begin{equation}
    y_p(t) = \frac{F_0}{m \sqrt{(\omega_0^2 - \omega^2)^2 + (\gamma \omega)^2}} \cos\left(\omega t - \phi\right),
\end{equation}
The steady-state solution is therefore:
\begin{equation}
    x(t) = x_0(\omega) \cos(\omega t - \phi),
\end{equation}
where
\begin{equation}
    x_0(\omega) = \frac{F_0}{m \sqrt{(\omega_0^2 - \omega^2)^2 + (\gamma \omega)^2}}.
\end{equation}
and
\[
\varphi(\omega) = \tan^{-1} \left( \frac{(b \omega / m)}{\omega_0^2 - \omega^2} \right)
\]

\subsection{Solution for the sound box system}
For the sound box system, we experimentally observe that the resonant frequency for the sound box is observed to be 300 Hz. The mass of the system, i.e., the sound box, is 256 \text{gm}.

We study the behavior of our model for $\gamma = 0\ s^{-1}$, $\gamma = 97.65\ {s^-1}$,$\gamma = 195.70\ s^{-1}$,$\gamma = 390.62\ s^{-1}$ and $\gamma = 585.93\ s^{-1}$.

\begin{figure}[htbp]
    \centering
    \begin{minipage}{0.45\textwidth}
        \centering
        \includegraphics[width=\textwidth]{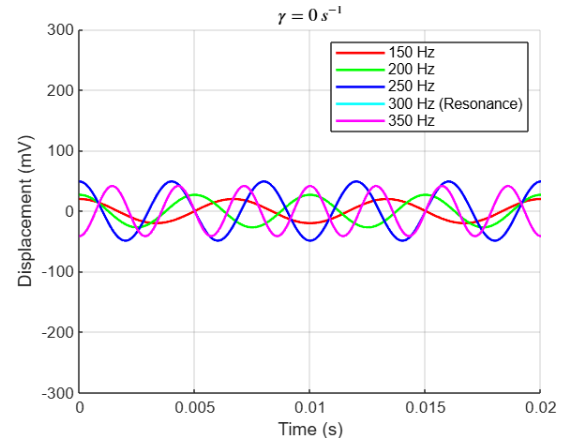}
        \caption*{(a)}
    \end{minipage}
    \hfill
    \begin{minipage}{0.45\textwidth}
        \centering
        \includegraphics[width=\textwidth]{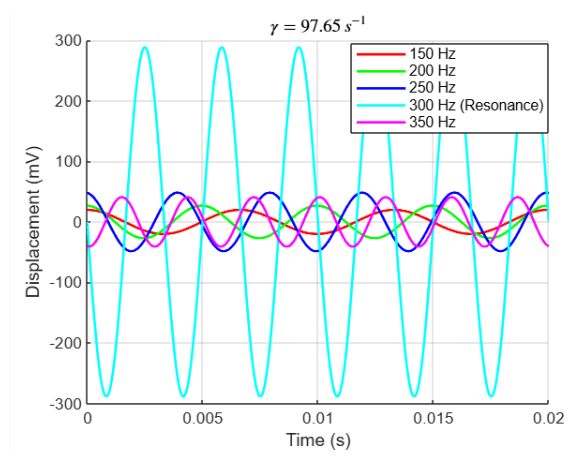}
        \caption*{(b)}
    \end{minipage}
    
    \vspace{0.5cm} 
    \begin{minipage}{0.45\textwidth}
        \centering
        \includegraphics[width=\textwidth]{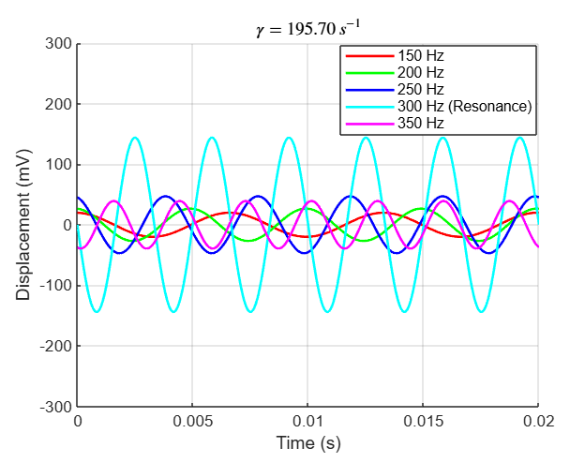}
        \caption*{(c)}
    \end{minipage}
    \hfill
    \begin{minipage}{0.45\textwidth}
        \centering
        \includegraphics[width=\textwidth]{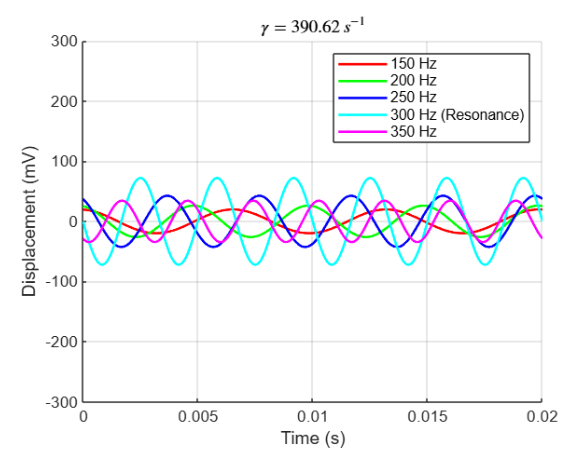}
        \caption*{(d)}
    \end{minipage}
    
    \vspace{0.5cm} 
    \begin{minipage}{0.45\textwidth}
        \centering
        \includegraphics[width=\textwidth]{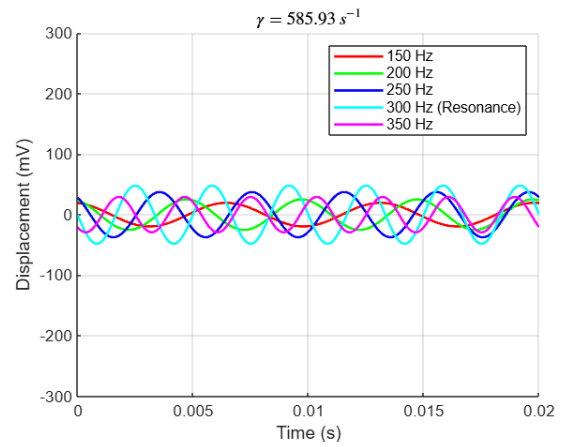}
        \caption*{(e)}
    \end{minipage}

    \caption{Steady state oscillation solutions to FHO for different values of $\gamma$. $\gamma= 195.70 s^-1$ provides the best theoretical fit to all experimental amplitudes for all frequencies.}
\end{figure}

We further observe that theoretically, the oscillation amplitudes achieved with varying $\gamma$ values are:

\begin{table}[H]
\centering

\begin{tabular}{|c|c|c|c|c|c|c|}
\hline
\textbf{Frequency} & \(\gamma = 0 \, \text{s}^{-1}\) & \(\gamma = 97.65 \, \text{s}^{-1}\) & \(\gamma = 195.70 \, \text{s}^{-1}\) & \(\gamma = 390.62 \, \text{s}^{-1}\) & \(\gamma = 585.93 \, \text{s}^{-1}\) \\ \hline
\(f_1 = 150 \, \text{Hz}\) & 39.86 & 39.84 & 39.76 & 39.48 & 39.09 \\ \hline
\(f_2 = 200 \, \text{Hz}\) & 53.81 & 53.71 & 53.40 & 52.22 & 50.42 \\ \hline
\(f_3 = 250 \, \text{Hz}\) & 97.84 & 96.87 & 94.13 & 85.17 & 74.63 \\ \hline
\(f_4 = 300 \, \text{Hz}\) & \(\infty\) & 577.02 & 287.93 & 144.25 & 96.17 \\ \hline
\(f_5 = 350 \, \text{Hz}\) & 82.78 & 81.65 & 78.49 & 68.79 & 58.41 \\ \hline
\end{tabular}

\caption{Theoretically predicted oscillation amplitudes (mV) for different frequencies at varying values of $\gamma$. $\gamma$ values characterize the response of the sound box to the forced oscillator and can be used to characterize different sound box materials.}
\end{table}

The resonant and near resonance frequencies are highly sensitive to changes in the $\gamma$ parameter. At $\gamma = 0$, 300 Hz is seen to have infinite oscillation amplitude. This can be readily derived;

The given equation is:
\begin{equation}
x_0(\omega) = \frac{F_0}{m \sqrt{(\omega_0^2 - \omega^2)^2 + (\gamma \omega)^2}}.
\end{equation}

For \( x_0(\omega) \to \infty \), the denominator must approach zero:
\begin{equation}
\sqrt{(\omega_0^2 - \omega^2)^2 + (\gamma \omega)^2} = 0.
\end{equation}

Squaring both sides, we get:
\begin{equation}
(\omega_0^2 - \omega^2)^2 + (\gamma \omega)^2 = 0.
\end{equation}

This equation is satisfied only when both terms individually equal zero:
\begin{equation}
(\omega_0^2 - \omega^2)^2 = 0 \quad \implies \quad \omega = \omega_0,
\end{equation}

\begin{equation}
(\gamma \omega)^2 = 0 \quad \implies \quad \gamma = 0.
\end{equation}

A good agreement for experimental and theoretical results is observed for; $\gamma=195.70\ s^{-1}$, m=0.256 kg and $F_0 = 13595666.87\  mV\ Kg $. 
\begin{figure}[H]
    \centering
    \includegraphics[width=0.8\linewidth]{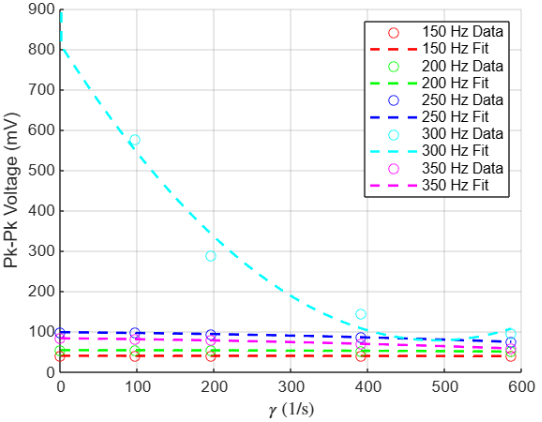}
    \caption{Peak to Peak voltage response of the sound box to different $\gamma$ values. The resonant and near to resonance frequency shows a high sensitive dependence on $\gamma$ values.}
\end{figure}

The sensitive dependence of the resonant frequency on the $\gamma$ can be readily seen mathematically.

The displacement amplitude \( x_0(\omega) \) of a damped harmonic oscillator under a periodic driving force is given by the expression:
\begin{equation}
x_0(\omega) = \frac{F_0}{m \sqrt{(\omega_0^2 - \omega^2)^2 + (\gamma \omega)^2}},
\end{equation}

At resonance, where the driving frequency matches the natural frequency of the oscillator (\(\omega = \omega_0\)), the amplitude reaches its maximum value. In this case, the term \((\omega_0^2 - \omega^2)^2\) in the denominator vanishes, and the amplitude reduces to:
\begin{equation}
x_0(\omega_0) = \frac{F_0}{m \gamma \omega_0}.
\end{equation}
This expression shows that the amplitude at resonance is inversely proportional to \(\gamma\). A small \(\gamma\) results in a sharp and large resonance peak, while an increase in \(\gamma\) causes the resonance peak to decrease in height and broaden. Thus, the amplitude at resonance is highly sensitive to changes in the damping parameter.

For frequencies near resonance, where \(|\omega - \omega_0| \ll \omega_0\), the term \((\omega_0^2 - \omega^2)^2\) remains small but nonzero, while the damping term \((\gamma \omega)^2\) is still significant. In this regime, both terms in the denominator contribute to the amplitude, and the sensitivity of the amplitude to changes in \(\gamma\) remains pronounced. A larger \(\gamma\) broadens the resonance peak and reduces the system's selectivity to frequencies near \(\omega_0\), while a smaller \(\gamma\) sharpens the resonance and increases the selectivity. The width of the resonance peak, often quantified as the full width at half maximum (FWHM), is directly proportional to \(\gamma\). This implies that the near-resonant response is governed by both the damping and the frequency deviation from \(\omega_0\), making it sensitive to \(\gamma\).

In contrast, at frequencies far from resonance, where \(|\omega - \omega_0| \gg \gamma\), the term \((\omega_0^2 - \omega^2)^2\) dominates the denominator of the amplitude expression. In this regime, the damping term \((\gamma \omega)^2\) becomes negligible compared to the frequency-dependent term \((\omega_0^2 - \omega^2)^2\). Consequently, the amplitude reduces to:

\begin{equation}
x_0(\omega) \approx \frac{F_0}{m (\omega_0^2 - \omega^2)},
\end{equation}

which depends primarily on the difference \((\omega_0^2 - \omega^2)\) and is largely unaffected by \(\gamma\). This indicates that the response of the system at frequencies far from resonance is insensitive to changes in the damping coefficient because the dominant factor is the large frequency mismatch rather than the damping.

\begin{figure}[H]
    \centering
    \begin{minipage}{0.45\textwidth}
        \centering
        \includegraphics[width=\linewidth]{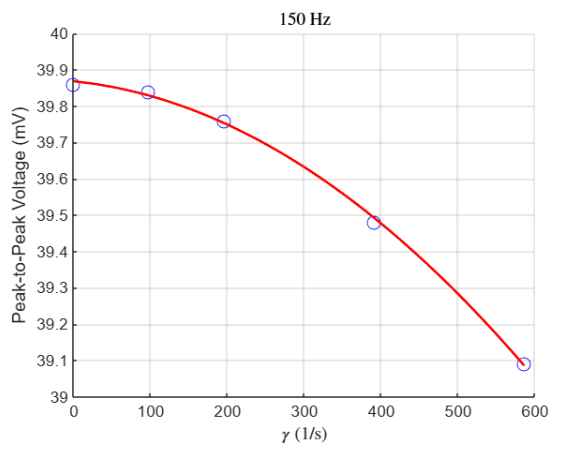}
        \caption*{(a) 150 Hz}
    \end{minipage}
    \hfill
    \begin{minipage}{0.45\textwidth}
        \centering
        \includegraphics[width=\linewidth]{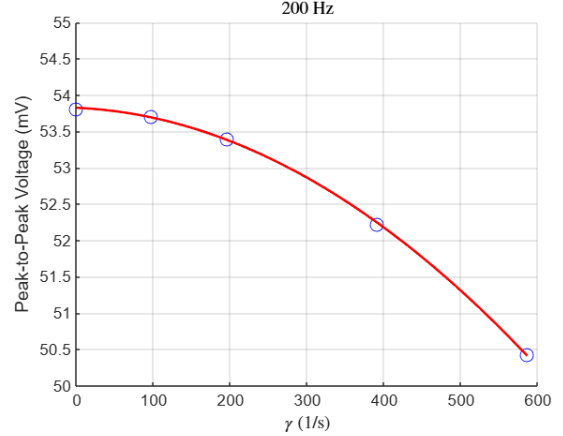}
        \caption*{(b) 200 Hz}
    \end{minipage}

    \vspace{1em}

    \begin{minipage}{0.45\textwidth}
        \centering
        \includegraphics[width=\linewidth]{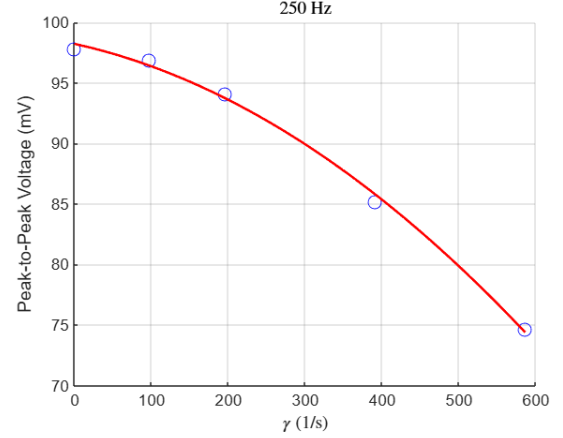}
        \caption*{(c) 250 Hz}
    \end{minipage}
    \hfill
    \begin{minipage}{0.45\textwidth}
        \centering
        \includegraphics[width=\linewidth]{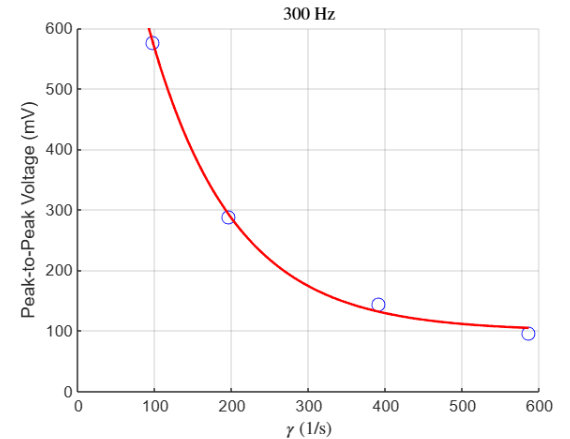}
        \caption*{(d) 300 Hz}
    \end{minipage}

    \vspace{1em}

    \begin{minipage}{0.5\textwidth}
        \centering
        \includegraphics[width=\linewidth]{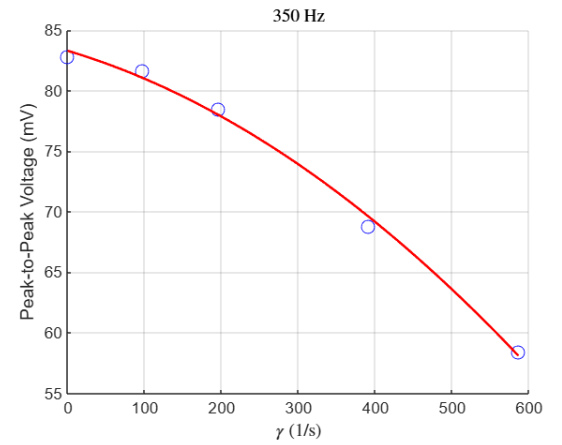}
        \caption*{(e) 350 Hz}
    \end{minipage}

    \caption{Peak-to-peak voltage as a function of $\gamma$ for various frequencies. The resonant frequency of 300 Hz has a sensitive dependence on the $\gamma$ value.}
\end{figure}

The sensitivity of the amplitude to the damping parameter \(\gamma\) depends on the proximity of the driving frequency to the natural frequency of the oscillator. At and near resonance, the amplitude is highly sensitive to changes in \(\gamma\), with \(\gamma\) directly influencing the sharpness and height of the resonance peak. However, far from resonance, the amplitude is governed by the frequency mismatch and the effect of \(\gamma\) becomes negligible.

\section{Conclusion}
The speaker setup demonstrates significant output intensity, particularly at resonant frequencies. The heat maps provide an effective insight into sound propagation and efficiency. These maps can be utilized to characterized sound propagation through different materials. The steady- state solution of FHO model was found to be in good agreement with the experimental results for the optimal parameter values of $\gamma=195.70\ s^{-1}$, m=0.256 kg and $F_0 = 13596\  V\ Kg $. This simple experiment provides valuable information about the workings of a speaker and a DIY manual to make your own speaker. We have demonstrated the sensitive dependence of near-resonant frequencies on damping parameter while operating in the damping-driven STL range.

\end{document}